\documentclass[apjl]{emulateapj}
\usepackage{amsmath} 
\usepackage{graphicx}
\usepackage{natbib}

\newcommand{\q}{\mathbf{q}}
\newcommand{\x}{\mathbf{x}}

\newcommand{\rr}{\mathbf{r}}
\newcommand{\avg}[1]{\langle{#1}\rangle}
\newcommand{\nablab}{\mathbf{\nabla}}

\begin{document}

\title{Removing BAO-peak shifts with local density transforms}
\author{Nuala McCullagh\altaffilmark{1}, Mark C. Neyrinck\altaffilmark{1}, Istv\'{a}n Szapudi\altaffilmark{2}, Alexander S. Szalay\altaffilmark{1}}
\altaffiltext{1}{Henry A. Rowland Department of Physics and Astronomy, The Johns Hopkins University 3400 N Charles St., Baltimore, MD 21218}
\altaffiltext{2}{Institute for Astronomy, University of Hawaii 2680 Woodlawn Drive, Honolulu, HI, 96822}

\begin{abstract}
Large-scale bulk flows in the Universe distort the initial density field, broadening the baryon-acoustic-oscillation (BAO) feature that was imprinted when baryons were strongly coupled to photons.  Additionally, there is a small shift inward in the peak of the conventional overdensity correlation function, a mass-weighted statistic.  This shift occurs when high density peaks move toward each other.  We explore whether this shift can be removed by applying to the density field a transform (such as a logarithm) that gives fairer statistical weight to fluctuations in underdense regions.  Using configuration-space perturbation theory in the Zel'dovich approximation, we find that the log-density correlation function shows a much smaller inward shift in the position of the BAO peak at low redshift than is seen in the overdensity correlation function.  We also show that if the initial, Lagrangian density of matter parcels could be estimated at their Eulerian positions, giving a displaced-initial-density field, its peak shift would be even smaller.  In fact, a transformed field that accentuates underdensities, such as the reciprocal of the density, pushes the peak the other way, outward. In our model, these shifts in the peak position can be attributed to shift terms, involving the derivative of the linear correlation function, that entirely vanish in this displaced-initial-density field.
\end{abstract}

\keywords{cosmology: theory -- large-scale structure of Universe}

\section{Introduction}

The ``standard ruler'' that baryon acoustic oscillations (BAO) provide is a major tool in cosmology.  The BAO peak has now been detected at unquestionable significance  \citep{eisensteinetal2005, cole2005, dr9bao2012}.  The power and cleanliness of the BAO method is such that despite these impressively large data sets, statistical uncertainties still dominate the systematic effects.  With the advent of even larger upcoming surveys, however, systematics such as the small shift in the BAO peak will soon be crucial to consider.

The peak is distorted and shifted at low redshift, even in the conceptually straightforward real-space dark-matter field, without galaxy bias and redshift-space distortions.  This happens because of bulk displacements, understood most easily in a Lagrangian description.  The broadening is largely removable by estimating the displacement field with the simple, linear Zel'dovich approximation \citep{EisensteinEtal2007, PadmanabhanEtal2012, mehta2011}.  Refinements in estimating the displacement field, such as using the log density, are also promising  \citep{mohayaee2006, lavaux2008,  FalckEtal2012, kitaura2012, TassevReconstruction2012}.
While restoring the initial sharpness of the BAO peak has obvious statistical benefits, such reconstruction methods can be rather involved.

As surveys grow, the shift in the peak becomes more important to consider in analysis than the broadening.  Various studies have attributed the shift to so-called ``mode-coupling" terms in the power spectrum \citep{smith2008, matsubaraRPT2008, seo2008, Padmanabhan2009, TaruyaNishimichi2009}.

We can also understand the peak shift in configuration space. The overdensity correlation function $\xi_\delta(r)=\avg{\delta(\x)\delta(\rr+\x)}_\x$ is a volume average of the overdensity $\delta=\rho/\bar{\rho}-1$.  The presence of the BAO peak in the linear correlation function means that on average in the initial conditions, over- and underdensities are surrounded by over- and underdense shells, respectively, at a radius $r=r_s$, the sound horizon.  In the initial conditions, the Gaussian probability density function (PDF) of $\delta_{\rm lin}$ exactly balances the contribution of the over-dense and under-dense regions to $\xi_\delta$.  

At low redshift, the PDF of $\delta$ becomes highly skewed, as overdensities collect matter and become sharp peaks (where $\delta$ grows arbitrarily high) at the expense of broad, shallow void regions (where $\delta\ge-1$).  If the BAO shell around each particle remained intact, tethered to move precisely with the particle toward or into clusters, the peak in $\xi_{\delta}$ would not shift from its original location.

However, 3D motions generally cannot preserve all BAO shells.  Central over- and underdensities attract or repel their  over- and underdense shells, broadening the averaged peak at low redshift.  Moreover, the overdense shells grow nonlinearly in density, and narrow (they must, from mass conservation); underdense shells broaden.   This nonlinear density enhancement of overdense shells at $\delta(\rr+\x)$ gives more weight to $\delta(\x)>0$ than $<0$, pulling the peak in $\xi(r)$ to slightly lower $r$.  This view of the peak shift is similar to that of  \citet{SherwinZaldarriaga2012}, in which the peak shifts because of an increased abundance of tracers in overdense regions, where the universe expands more slowly than average.

In this Letter, we use the Zel'dovich approximation \citep{zeldovich1970} to test the hypothesis that a density transform that produces a more-Gaussian final-conditions density PDF leads to a reduction in the peak motion, as such a transform more equally weights the contributions in the volume average from overdense and underdense regions. While we do not show mathematically how transformations change the weightings in the correlation function average, which would involve manipulating the PDFs of the transformed fields, we verify that the more weight low densities receive in a transformation, the larger the peak position. We investigate two transformations that produce near-Gaussian PDFs --  the log density and the displaced-initial-density -- and one that over-weights low densities, $(1+\delta)^{-1}$.

Our investigation of log-density statistics is also motivated by recent work on clustering statistics of the log density and Gaussianized density, which have dramatically lower covariance on small scales, as well as better fidelity to the linear-theory shape \citep{NeyrinckEtal2009}.  In principle, they can then give much tighter constraints on cosmological parameters \citep{Neyrinck2011}.  In fact, analysis of the log-density correlation function accesses all of the Fisher information in an idealized entirely lognormal field, whereas using even arbitrarily high-point  $\delta$ correlation functions can give only a small fraction of that total information \citep{Carron2011,CarronNeyrinck2012}.  Here, for the first time we explicitly show the benefits of a log transform on large scales, where the benefits might have been thought to be negligible.

\section{Theory}

In this section, we derive the nonlinear correlation functions of the Zel'dovich density, the log density, and the displaced-initial-density field. We follow the approach outlined by \citet{mccullaghszalay2012} for perturbatively calculating the nonlinear density correlation function in the Zel'dovich approximation. We briefly review this approach below, and extend it to calculate the nonlinear correlation functions of the log density and the Lagrangian density $\delta_L$ at an Eulerian position $\x$. For the latter, we examine the effect of Zel'dovich displacements by themselves, without considering the change in density at a constant Lagrangian coordinate.

We start with the Zel'dovich approximation, which maps particles' initial Lagrangian coordinate, $\q$, to their co-moving Eulerian coordinate, $\x$, through the gradient of the initial displacement potential, $\phi(\q)$, and the growth function $D(t)$ \citep{zeldovich1970}.  The initial displacement potential is related to the initial density field through the Poisson equation:
\begin{align}
\x(\q, t) &= \q-D(t) \nablab_q \phi(\q)\ ,\label{eq:Zel'dovich}\\
 D(t)\nabla_q^2\phi(\q)&=\delta_{0}(\q,t)\ .\notag
\end{align}
The density can be written in terms of the Jacobian of the transformation from Lagrangian to Eulerian coordinates from the conservation of mass.
 \begin{align}
\frac{\rho(\x, t)}{\bar \rho}&=\left | \frac{\partial x_i}{\partial q_j} \right |^{-1}=\frac{1}{J(\q, t)} =  1 + \delta(\q(\x))\ .
\label{eq:euleriandensity}
\end{align}
Here, $\delta$ is the nonlinear overdensity as a function of Lagrangian position. Equation (\ref{eq:euleriandensity}) for the Eulerian density is only strictly valid before shell-crossing, where the mapping from $\q$ to $\x$ is one-to-one. 

The Jacobian can be written in terms of the eigenvalues ($\lambda_1$, $\lambda_2$, and $\lambda_3$) of the symmetric matrix $d_{ij}(\q)$ \citep{zeldovich1970}, or equivalently in terms of the invariants of the matrix ($I_1$, $I_2$, and $I_3$), which are simple functions of the eigenvalues \citep{bouchet1995}.
\begin{align}
d_{ij}(\q)&=\frac{\partial^2 \phi(\q)}{\partial q_i \partial q_j}\ ,\\
J(\q, t)&=(1-D\lambda_1)(1-D\lambda_2)(1-D\lambda_3)\label{eq:jaceigen}\\
&=1-D I_1(\q)+D^2 I_2(\q)-D^3 I_3(\q)\ .
\end{align}

The nonlinear overdensity can thus be expressed in terms of the initial quantities by a Taylor expansion of the inverse Jacobian, to any order:
\begin{align}
	\delta(\q,t)&=DI_1(\q)+D^2\left(I_1(\q)^2-I_2(\q)\right)\notag\\
	&\qquad+D^3\left(I_1(\q)^3-2I_1(\q)I_2(\q)+I_3(\q)\right)+...
	\label{eq:lagrangedensity}
\end{align}

In order to express the overdensity as a function of final (Eulerian) position, $\x$, we Taylor expand Equation (\ref{eq:lagrangedensity}) about the point $\x=\q$. The nonlinear density as a function of $\x$ to third order in $D$ is then:
\begin{align}
	\delta(\x,t)&=\bigg(\delta(\q,t) + D \sum_i \frac{\partial\phi(\q)}{\partial q_i}\frac{\partial\delta(\q,t)}{\partial q_i}\notag\\
	&+D^2\sum_{i,j}\frac{\partial^2 \phi(\q)}{\partial q_i\partial q_j}\frac{\partial \phi(\q)}{\partial q_j}\frac{\partial \delta(\q,t)}{\partial q_i}\notag\\
	&\qquad+ \frac{1}{2}D^2 \sum_{i,j}\frac{\partial^2\delta(\q,t)}{\partial q_i \partial q_j}\frac{\partial \phi(\q)}{\partial q_i}\frac{\partial \phi(\q)}{\partial q_j}\bigg)\Bigg|_{\q=\x}\ .
	\label{eq:eulerdensity}
\end{align}

The correlation function in co-moving Eulerian coordinates can be written in powers of $D$ using the above expansion of the Eulerian overdensity in terms of the initial quantities. Because the initial overdensity is assumed to be a zero-mean Gaussian random field, the odd moments vanish. The first two terms of the correlation function are then:
\begin{align}
\xi_{\delta}(\rr, t)&\equiv\langle \delta(\x, t)\delta(\x+\rr, t)\rangle=\xi_{\delta}^{(1)}(\rr)D^2+\xi_{\delta}^{(2)}(\rr)D^4+...
\end{align}

We define the functions:
\begin{align}
\xi_n^m(r)&=\frac{1}{2\pi^2} \int_0^{\infty} \!P_{\mathrm{lin}}(k)j_n(kr)k^{m+2}\mathrm{d}k\label{xinm}\text{ ,}
\end{align}
where $j_n$ is the spherical Bessel function of order $n$ and $P_{\mathrm{lin}}(k)$ is the power spectrum of the initial density fluctuations. 
Using this definition, the linear term in the expansion of the correlation function is:
\begin{align}
\xi_{\delta}^{(1)}(\rr)&=\xi_0^0(r)\text{ ,}
\end{align}
the spherically symmetric Fourier transform of the linear power spectrum. 

The first nonlinear term in the correlation function can be calculated using spherical harmonics. For more details on the calculation, see \citet{mccullaghszalay2012}. The final expression is:
\begin{align}
	\xi_{\delta}^{(2)}(\rr)&= \frac{19}{15}\xi_0^0(r)^2+\frac{34}{21}\xi_2^0(r)^2\notag
	+\frac{4}{35}\xi_4^0(r)^2 - \frac{16}{5}\xi_1^{-1}(r)\xi_1^1(r)\\
	&-\frac{4}{5}\xi_3^{-1}(r)\xi_3^1(r)+\frac{1}{3}\xi_0^{-2}(r)\xi_0^2(r)-\frac{1}{3}\xi_0^{-2}(0)\xi_0^2(r)
	\notag\\
	&+\frac{2}{3}\xi_2^{-2}(r)\xi_2^2(r) 
\label{eq:nlcf}
\end{align}

We can use the same approach to calculate the correlation function of the log density field. Starting with Equation (\ref{eq:euleriandensity}) and using the definition of the Jacobian in Equation (\ref{eq:jaceigen}), we write the log of the density, which we define as the quantity $A(\q)$, as:
\begin{align}
\ln\left(1+\delta(\q, t)\right)&\equiv A(\q)=-\ln J(\q, t)\notag\\
&=-(\ln(1- D \lambda_1)+\ln(1-D\lambda_2)\notag\\
&\qquad+\ln(1-D\lambda_3))\ .
\end{align}

We expand this expression in powers of $D$, and rewrite it in terms of the invariants of the deformation tensor:
\begin{align}
\ln\left(1+\delta(\q, t)\right)&=D(\lambda_1+\lambda_2+\lambda_3)+\frac{1}{2}D^2(\lambda_1^2+\lambda_2^2+\lambda_3^2)\notag\\
&\qquad+\frac{1}{3}D^3(\lambda_1^3+\lambda_2^3+\lambda_3^3)+...\\
&=D I_1+\frac{1}{2}D^2(I_1^2-2I_2)\notag\\
&\qquad+\frac{1}{3}D^3(I_1^3-3I_1I_2+3I_3)+...
\end{align}

We now have the equivalent of Equation (\ref{eq:lagrangedensity}), but for the log density field. We transform this into Eulerian coordinates as above to get an expression for $A(\x, t)$. Because this quantity has a non-zero mean, we define the correlation function as:
\begin{align}
\xi_A(\rr, t)&=\left\langle (A(\x, t)-\bar{A}) (A(\x+\rr, t)-\bar{A})\right\rangle\ \text{,}
\end{align}
where $\bar{A}$ is the Eulerian mean.

The first two terms in the correlation function are then:
\begin{align}
\xi_A^{(1)}(r)&=\xi_0^0(r)\\
\xi_A^{(2)}(r)&=-\frac{2}{3}\xi_0^0(0)\xi_0^0(r)+ \frac{13}{30}\xi_0^0(r)^2+\frac{20}{21}\xi_2^0(r)^2+\frac{4}{35}\xi_4^0(r)^2 \notag\\
&\qquad
	- \frac{6}{5}\xi_1^{-1}(r)\xi_1^1(r)-\frac{4}{5}\xi_3^{-1}(r)\xi_3^1(r)+\frac{1}{3}\xi_0^{-2}(r)\xi_0^2(r)\notag\\
	&\qquad-\frac{1}{3}\xi_0^{-2}(0)\xi_0^2(r)
	+\frac{2}{3}\xi_2^{-2}(r)\xi_2^2(r) 
\end{align}

Note that the first term in $\xi_A^{(2)}(r)$ can be written as $-\frac{2}{3}\sigma_0^2 \xi_A^{(1)}(r)$, where $\sigma_0^2$ is the variance of the initial density field (where the density is assumed to be smoothed on some scale). This term describes the reduction in the amplitude of $\xi_A$ compared with the linear correlation function.

Next, we look at the initial density field at Eulerian position $\x$. For this quantity, we take the initial density, $\delta_0(\q)$, and use the Zel'dovich formula, Equation (\ref{eq:Zel'dovich}), to relate $\q$ to $\x$. The expression for $\delta_L(\x)$ (where the subscript $L$ stands for Lagrangian) is equivalent to Equation (\ref{eq:eulerdensity}), but with $\delta$ replaced with $\delta_0$:
\begin{align}
\delta_L(\x)&=\Bigg(\delta_0(\q) + D^2 \sum_i \frac{\partial \phi(\q)}{\partial q_i}\frac{\partial \delta_0(\q)}{\partial q_i} \notag\\
&\qquad+ D^3 \sum_{i,j}\frac{\partial^2 \phi(\q)}{\partial q_i \partial q_j} \frac{\partial \phi(\q)}{\partial q_j}\frac{\partial \delta_{0}(\q)}{\partial q_i}\notag\\
&\qquad+\frac{D^3}{2}\sum_{i,j} \frac{\partial^2 \delta_{0}(\q)}{\partial q_i \partial q_j}\frac{\partial \phi(\q)}{\partial q_i}\frac{\partial \phi(\q)}{\partial q_j}\Bigg)\Bigg|_{\q = \x}
\end{align}

We label the correlation function of this quantity as $\xi_{\delta_L}$. The first two terms are:
\begin{align}
\xi_{\delta_L}^{(1)}(r)&=\xi_0^0(r)\\
\xi_{\delta_L}^{(2)}(r)&=-\frac{2}{3}\xi_0^0(0)\xi_0^0(r)+\frac{1}{3}\xi_0^0(r)^2+\frac{2}{3}\xi_2^0(r)^2+\frac{1}{3}\xi_0^{-2}(r)\xi_0^2(r)\notag\\
&\qquad-\frac{1}{3}\xi_0^{-2}(0)\xi_0^2(r)+\frac{2}{3}\xi_2^{-2}(r)\xi_2^2(r)
\end{align}

Again, we note the damping term, $-\frac{2}{3}\xi_0^0(0)\xi_0^0(r)$, which is the same as we found in $\xi_A$. We also note that the expression for $\xi_{\delta_L}^{(2)}$ is simpler than either $\xi_{\delta}^{(2)}$ or $\xi_{A}^{(2)}$.

Finally, to investigate the peak shift in a transformation which further boosts the weight of underdensities, we consider the correlation function of the reciprocal of the density, $\xi_{1/\rho}$. This statistic is of some theoretical interest in a Lagrangian approach, since $(1+\delta)^{-1}$ is simply the Jacobian, Equation (\ref{eq:jaceigen}). When we expand to Eulerian coordinates we get:
\begin{align}
(1+\delta(\x))^{-1}&=\Bigg(J(\q) + D^2 \sum_i \frac{\partial \phi(\q)}{\partial q_i}\frac{\partial J(\q)}{\partial q_i} \notag\\
&\qquad+ D^3 \sum_{i,j}\frac{\partial^2 \phi(\q)}{\partial q_i \partial q_j} \frac{\partial \phi(\q)}{\partial q_j}\frac{\partial J(\q)}{\partial q_i}\notag\\
&\qquad+\frac{D^3}{2}\sum_{i,j} \frac{\partial^2 J(\q)}{\partial q_i \partial q_j}\frac{\partial \phi(\q)}{\partial q_i}\frac{\partial \phi(\q)}{\partial q_j}\Bigg)\Bigg|_{\q = \x}
\end{align}

The first two terms of $\xi_{1/\rho}$ are:
\begin{align}
\xi_{1/\rho}^{(1)}(r)&=\xi_0^0(r)\\
\xi_{1/\rho}^{(2)}(r)&=\frac{2}{3}\xi_0^0(0)\xi_0^0(r)+ \frac{3}{5}\xi_0^0(r)^2+\frac{2}{7}\xi_2^0(r)^2+\frac{4}{35}\xi_4^0(r)^2 \notag\\
&\qquad
	+ \frac{4}{5}\xi_1^{-1}(r)\xi_1^1(r)-\frac{4}{5}\xi_3^{-1}(r)\xi_3^1(r)+\frac{1}{3}\xi_0^{-2}(r)\xi_0^2(r)\notag\\
	&\qquad-\frac{1}{3}\xi_0^{-2}(0)\xi_0^2(r)
	+\frac{2}{3}\xi_2^{-2}(r)\xi_2^2(r) 
\end{align}

Because our interest in this transformation is to test a more extreme weighting of underdensities, and we do not anticipate a practical measurement of $\xi_{1/\rho}$ (which, for example, could be violently sensitive to discreteness noise), for clarity we do not plot $\xi_{1/\rho}$ in the figures below, but do discuss its peak position.

\section{Numerical Results}
Our goal is to understand how the various transformations to the density field affect the correlation function, specifically at the BAO peak. In this section, we consider the effects of the first nonlinear term in the correlation functions on the BAO peak position and shape. We then examine the various terms in the expressions in order to understand their role in the overall correlation function. For the following we use an initial power spectrum generated by {\scshape camb} \citep{camb}, assuming fiducial WMAP5 cosmological parameters \citep{wmap5}: ($H_0$, $\Omega_{\rm \Lambda}$, $\Omega_{\rm CDM}$, $\Omega_b$, $n_s$, $\sigma_8$) = ($70.1, 0.233, 0.0462, 0.721, 0.96, 0.817$). We smooth the linear power spectrum with a $\sigma=5\ \mathrm{Mpc}/h$ Gaussian, which was a large enough smoothing length to give agreement between the analytical expression and numerical Zel'dovich realizations in \citet{mccullaghszalay2012} up to $z=0$.

\begin{figure}[h]
\begin{center}
\includegraphics[scale=0.34]{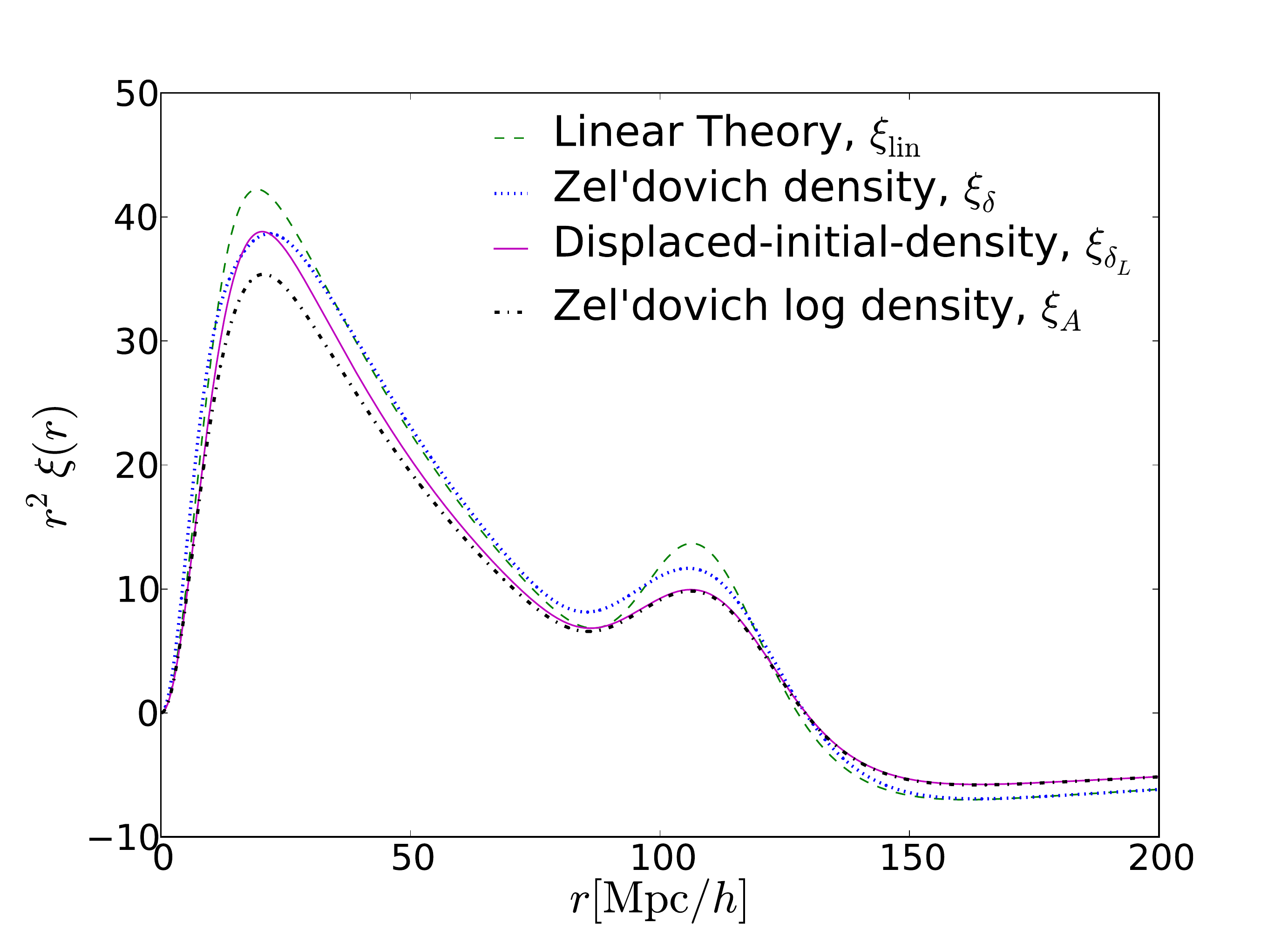}
\caption{Nonlinear correlation functions at $z=0$. Green dashed line - linear theory, Blue dotted line - nonlinear density from Zel'dovich approximation, magenta solid line - displaced-initial-density field, $\delta_L(\vec x)$, black dot-dashed line - $\ln(1+\delta)$ from Zel'dovich approximation }
\label{fig:cf_all}
\end{center}
\end{figure}

Fig.\ \ref{fig:cf_all} shows the correlation functions at $z=0$ for the linear theory density, Zel'dovich density, Zel'dovich log-density, and displaced-initial-density. We note that the BAO peak of all nonlinear correlation functions are broadened compared with $\xi_{\mathrm{lin}}$. The smoothing of the acoustic peak relative to that of the linear correlation function is due to large-scale bulk motions, which are present in the three nonlinear fields. In the case of $\xi_{\delta}$ we also observe an enhancement at the peak, perhaps from over-weighting overdense mass elements, which contract.

\begin{figure}[h]
\includegraphics[scale=0.34]{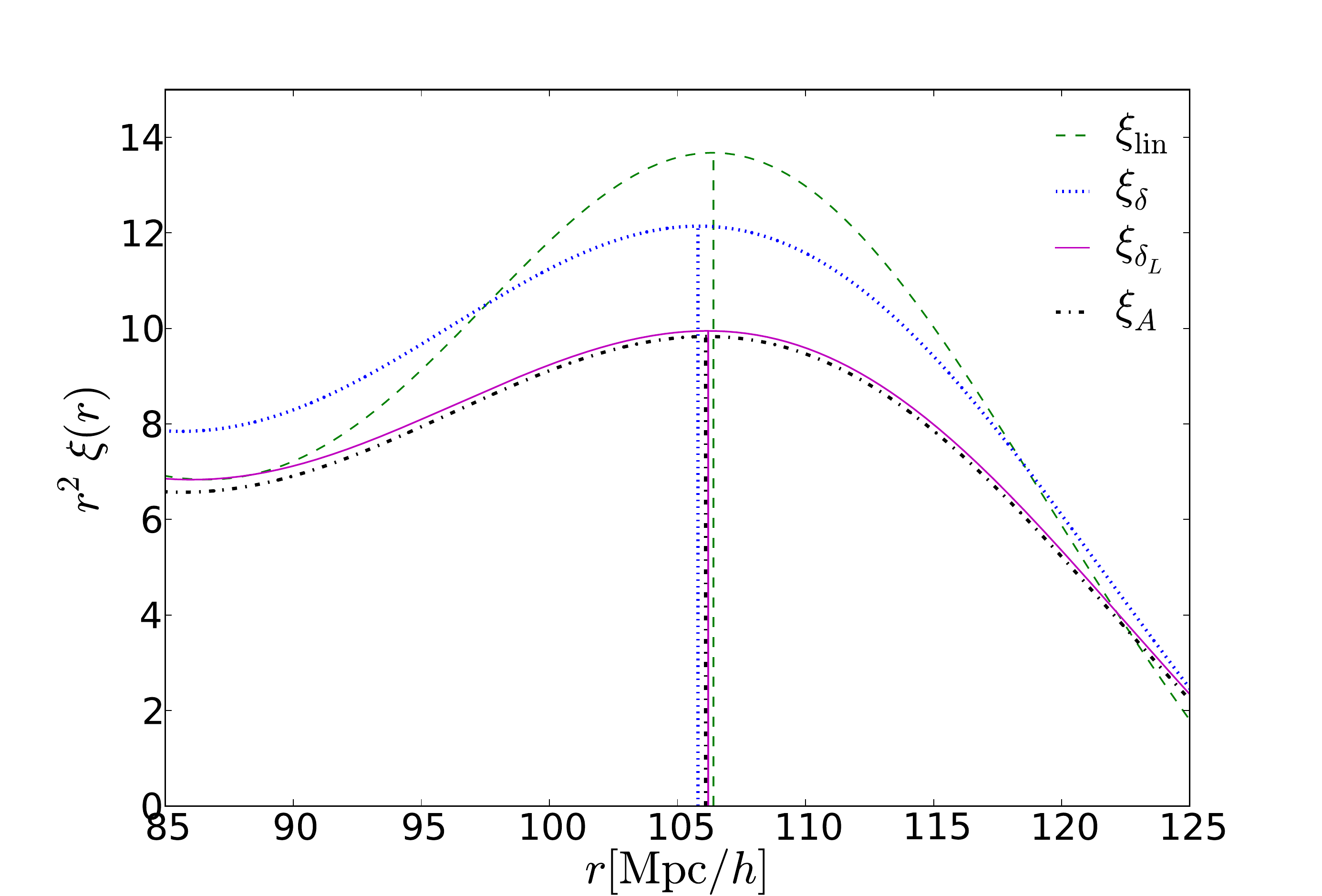}
\caption{Peak locations of the various correlation functions at $z=0$. Green dashed line - linear theory, peak is at $r=106.4 \mathrm{Mpc}/h$. Blue dotted line - nonlinear density from Zel'dovich approximation, peak is at $r=105.8\ \mathrm{Mpc}/h$. Magenta solid line - displaced-initial-density, $\delta_L(\vec x)$, peak is at $r=106.2\ \mathrm{Mpc}/h$. Black dot-dashed line - $\ln(1+\delta)$ from Zel'dovich approximation, peak is at $r=106.1\ \mathrm{Mpc}/h$. }
\label{fig:peak_positions}
\end{figure}

Fig.\ \ref{fig:peak_positions} shows a close-up of the BAO peak, and the position of each of the peaks ($r_p$) is indicated. We measure the peak of $r^2 \xi$ because this is the quantity that is integrated to give the variance on a given scale. The peaks in order of largest to smallest $r_p$ are: $\xi_{1/\rho}: r_p=106.5\ \mathrm{Mpc}/h$ (not shown), $\xi_{\mathrm{lin}}: r_p=106.4\ \mathrm{Mpc}/h$, $\xi_{\delta_L}: r_p=106.2\ \mathrm{Mpc}/h$, $\xi_A: r_p=106.1\ \mathrm{Mpc}/h$, and $\xi_{\delta}: r_p=105.8\ \mathrm{Mpc}/h$. There are other roughly equivalent definitions of the peak position, such as the minimum $\chi^2$ of template correlation functions or the peak through a wavelet filter. These are essential to use in the presence of cosmic variance, but the simpler peak definition used here is acceptable since we use a linear correlation function with no cosmic variance.

\begin{figure}[h]
\includegraphics[scale=0.42]{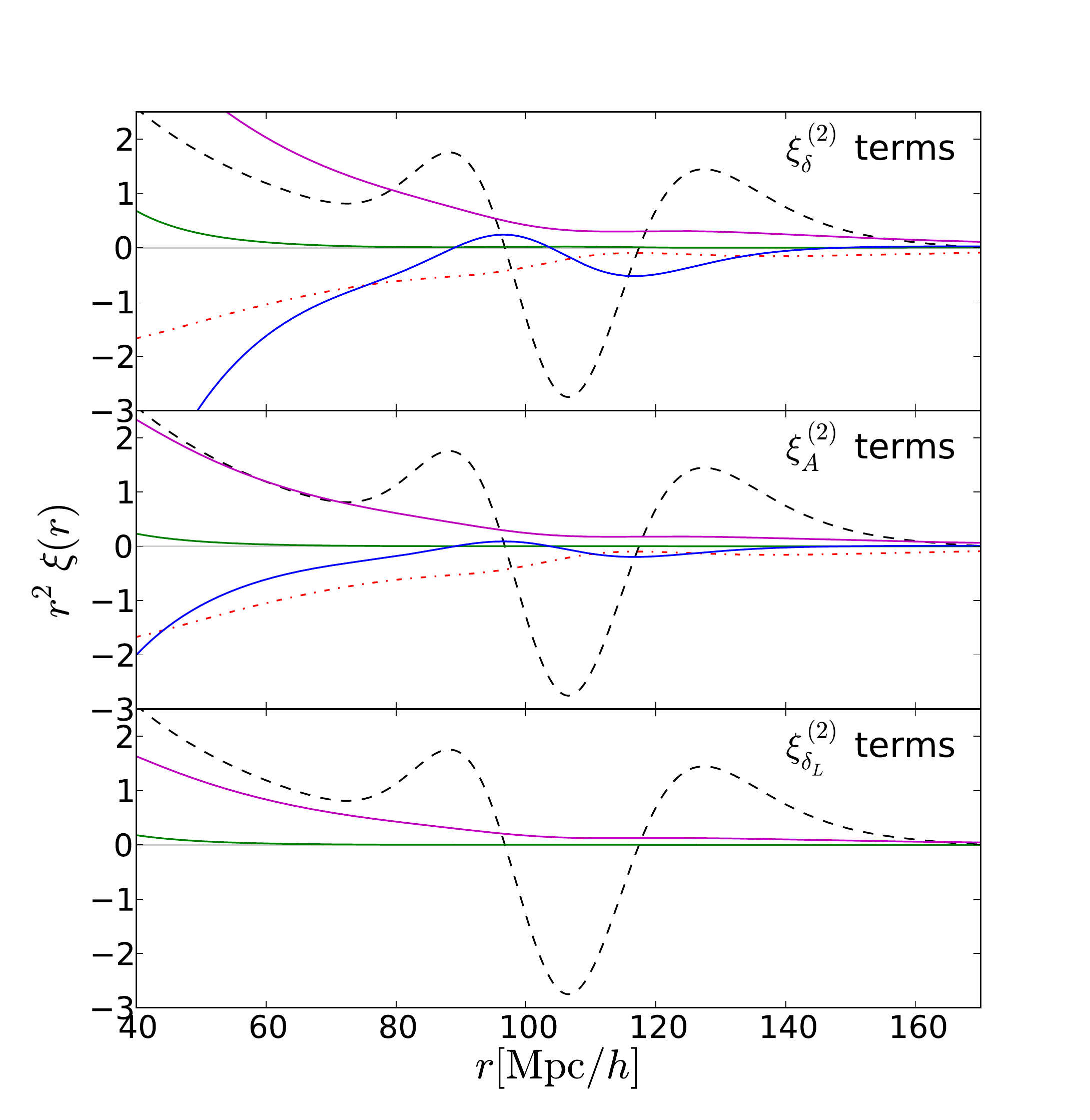}
\caption{Top to bottom: Terms in the expressions for $\xi_{\delta}^{(2)}$,  $\xi_{A}^{(2)}$, and  $\xi_{\delta_L}^{(2)}$ (not including damping terms in the last two). Different terms are represented by different colors: Black dashed - terms with same amplitude in all 3 expressions ($1/3(\xi_0^2(r)\xi_0^{-2}(r)-\xi_0^2(r)\xi_0^{-2}(r)) + 2/3 \xi_2^2(r)\xi_2^{-2}(r)$). Red dot-dash - terms that are the same in both $\xi_{\delta}$ and $\xi_A$ ($-4/5 \xi_3^1(r)\xi_3^{-1}(r) + 4/35 \xi_4^0(r)^2$). The amplitude of this term is zero for $\xi_{\delta_L}$. Green - amplitude of $\xi_0^0(r)^2$ term. Magenta - amplitude of $\xi_2^0(r)^2$ term. Blue - amplitude of $\xi_1^1(r)\xi_1^{-1}(r)$.
}
\label{fig:cf_terms}
\end{figure}

Next, we look at the effects of individual terms in the nonlinear expressions on the location of the peak. Fig.\ \ref{fig:cf_terms} shows the terms that contribute to each of these nonlinear correlation functions. We exclude the damping terms of $\xi_A$ and $\xi_{\delta_L}$ because they do not contribute to the shift in the peak position. Each term is represented by a different color, and the solid lines have a different amplitude in each expression. The dot-dashed line has the same amplitude in $\xi_{\delta}$ and $\xi_A$ (and is zero in $\xi_{\delta_L}$), and the dashed line is the same in all three expressions. From this plot we see that the effect of the dashed black line is to smooth the peak in all three cases, by increasing the amplitude around the peak and decreasing the amplitude at the peak. Several of the other terms cause a shift in the peak position by showing a slope around the peak. For $\xi_{\delta_L}$, the magenta line is the only term with this property, so the shift is not very large. In $\xi_{\delta}$ and $\xi_A$, the red (dot-dashed) and magenta (solid) lines have nearly opposite slopes, and so the shift from the combination of these terms is small. 

The major contributor to the shift in the peak position in $\xi_A$ and $\xi_{\delta}$ is the (solid) blue line, which is the term proportional to $-\xi_1^1(r)\xi_1^{-1}(r)$. In $\xi_{\delta}$ the amplitude of this term is $16/5$, whereas for $\xi_A$ the amplitude is $6/5$. Thus we see a larger shift in the peak position in $\xi_{\delta}$ than in $\xi_A$, and both have larger shifts than $\xi_{\delta_L}$. We also note that in the expression for $\xi_{1/\rho}$ this term has the opposite sign, indicating a shift to greater radius.

We can understand this term better by using the recursion relation for spherical Bessel functions:
\begin{align}
j_0'(x)&=-j_1(x)
\end{align}
This allows us to rewrite $\xi_1^m(r)$ in terms of derivatives of the linear correlation functions with respect to $r$:
\begin{align}
\xi_1^m(r)&=-\xi_0^{m-1}(r)'\\
\xi_1^1(r)\xi_1^{-1}(r)&=\xi_0^0(r)'\xi_0^{-2}(r)'
\end{align}
In fact this is exactly the term \citet{crocce2008} found to be responsible for the BAO peak shift from the mode coupling term of the correlation function in renormalized perturbation theory. This term is the product of the derivative of the linear density correlation function ($\xi_0^0(r)$) and the derivative of the linear velocity correlation function ($\xi_0^{-2}(r)$).

\section{Conclusion}

In this Letter, we have explored the idea that the shift in the position of the BAO peak in the usual density correlation function is due to over-weighting of high density peaks that bulk flows bring toward each other.  We have shown that the two transformations on the density field that suppress overdense regions and boost underdense regions in statistical weight result in a reduced shift in the BAO peak position at low redshift. Moreover, we have shown that a transformation that further weights underdense regions causes the peak to shift in the opposite direction, to larger $r$. We used the Zel'dovich approximation and configuration-space perturbation theory to examine the first nonlinear contribution to the correlation functions of the nonlinear density, the log-density, and the displaced-initial-density fields. We identified the terms largely responsible for the shift in the BAO peak, and found that they were proportional to the first derivative of the linear correlation function. The displaced-initial-density exhibits the smallest shift due to a lack of shift terms in the nonlinear correlation function expression. 

The log-density correlation function has some shift, but it is smaller than in the nonlinear density correlation function. In fact, the correlation function of the log density is almost indistinguishable from that of the displaced-initial-density on BAO scales.  The near-equivalence of the log density and the displaced initial density recalls the lognormal model of \citet{ColesJones1991}, in which gravity enhances initial densities with an exponential transform, even though a Zel'dovich-realization density field is not an exactly lognormal field.

By studying both the nonlinear density and displaced-initial-density, we decouple the effects of density enhancements from the bulk flows of initial fluctuations. The fundamental broadening of the initial peak is seen in $\xi_{\delta_L}$, whereas in $\xi_{\delta}$ the peak is additionally sharpened. However, we suspect that this sharpness is not statistically useful because it comes from only the fraction of the mass elements that have contracted. We intend to test the noise properties of the transformed fields in $N$-body simulations. Future work is also needed to examine the potential of transformations such as these in the presence of galaxy bias, discreteness, and redshift-space distortions.

\acknowledgements

We thank Xin Wang and Donghui Jeong for many useful discussions, and the referee for a helpful report. MN, AS, and NM acknowledge the Gordon and Betty Moore Foundation and NSF OIA grant CDI-1124403. MN acknowledges a New Frontiers grant from the John Templeton Foundation. IS acknowledges NASA grants NNX12AF83G and NNX10AD53G and the Polanyi program of the Hungarian National Office for the Research and Technology (NKTH).

\bibliographystyle{apj}
\bibliography{logzabib}

\end{document}